# Observation of coupling between zero- and two-dimensional semiconductor systems based on anomalous diamagnetic effects

Shuo Cao[1,§], Jing Tang[1,§], Yue Sun[1], Kai Peng[1], Yunan Gao[1], Yanhui Zhao[1], Chenjiang Qian[1], Sibai Sun[1], Hassan Ali[1], Yuting Shao[1], Shiyao Wu[1], Feilong Song[1], David A. Williams[2], Weidong Sheng[3], Kuijuan Jin[1,4], and Xiulai Xu[1]

[1]Beijing National Laboratory for Condensed Matter Physics, Institute of Physics, Chinese Academy of Sciences, Beijing 100190, China
[2] Hitachi Cambridge Laboratory, Cavendish Laboratory, Cambridge CB3 0HE, United Kingdom
[3] Department of Physics, Fudan University, Shanghai 200433, China
[4] Collaborative Innovation Center of Quantum Matter, Beijing 100084, China
[§]These authors contributed equally to this work.

Address correspondence to Xiulai Xu , xlxu@iphy.ac.cn; Kuijuan Jin, kjjin@iphy.ac.cn.

**We report the direct observation of coupling between a single self-assembled InAs quantum dot and a wetting layer, based on strong diamagnetic shifts of many-body exciton states using magneto-photoluminescence spectroscopy. An extremely large positive diamagnetic coefficient is observed when an electron in the wetting layer combines with a hole in the quantum dot; the coefficient is nearly one order of magnitude larger than that of the exciton states confined in the quantum dots. Recombination of electrons with holes in a quantum dot of the coupled system leads to an unusual negative diamagnetic effect, which is five times stronger than that in a pure quantum dot system. This effect can be attributed to the expansion of the wavefunction of remaining electrons in the wetting layer or the spread of electrons in the excited states of the quantum dot to the wetting layer after recombination. In this case, the wavefunction extent of the final states in the quantum dot plane is much larger than that of the initial states because of the absence of holes in the quantum dot to attract electrons. The properties of emitted photons that depend on the large electron wavefunction extents in the wetting layer indicate that the coupling occurs between systems of different dimensionality, which is also verified from the results obtained by applying a magnetic field in different configurations. This study paves a new way to observe hybrid states with zero- and two-**

**dimensional structures, which could be useful for investigating the Kondo physics and implementing spin-based solid-state quantum information processing.**

Semiconductor quantum dots (QDs), also known as "artificial atoms", have attracted considerable interest because of their application to quantum optoelectronic devices such as single-photon sources [1–6], quantum bits [7, 8], quantum logic gates [9], and photon-spin interfaces [10–12]. Owing to the quasi-zero-dimensional nature of QDs, many-body effect induced exciton states have been achieved in single QDs by controlling charge states with external electric and magnetic fields [13–15]. Following the discovery of graphene [16], two-dimensional materials, such as $MoS_2$ [17, 18] and black phosphorous [19], have been investigated extensively. Recently, the coupling between QDs and two-dimensional extended states has been investigated to demonstrate the Kondo physics [15, 20–24] and Fano effect [25, 26]. Hybridization between a single QD and a two-dimensional continuum state has been studied experimentally and theoretically in various systems [21, 27–32]. For self-assembled QDs grown by molecular beam epitaxy (known as Stranski–Krastanov growth), a wetting layer with a thickness of a few monolayers is always formed underneath the QDs [33–35], which naturally provides a platform for the formation of coupled systems for many-body exciton states [15, 27]. Karrai et al. [27] have shown the coherent hybridization of localized QD states and extended continuum states in the wetting layer; specifically, the triply negatively charged exciton states in the QD coherently couple with the Landau levels of the wetting layer in a magnetic field. Usually, in the investigation of the optical properties of single QDs, pumping power is minimized to avoid mixed emission from neighboring QDs and simplify the explanation of spectra [36]. However, coupling of excited states of higher negative charges in QDs with a wetting layer, which for instance, can be achieved with a high excitation power, has not been reported.

Magneto-photoluminescence spectroscopy of single QDs has been used extensively to study the exciton wavefunction distribution through the diamagnetic effect [37–40]. In a weak magnetic field, the exciton diamagnetic shift reflects the spatial wavefunction distribution and Coulomb interactions in QDs [38]. Usually, the magnetic field provides an additional confinement in the QD plane, thus increasing the ground state energy with a diamagnetic coefficient of approximately 10.0 μeV/T$^2$ for single InAs QDs [39, 40]. The increase of ground state energy with increasing magnetic field induces a "positive" diamagnetic shift. Recently, "negative" diamagnetic shifts have also been observed in negatively charged exciton recombination because of different wavefunction distributions of the initial and the final states in single QDs [39–41]. Because of the different confinements of carriers in zero-dimensional quantum dots and two-dimensional wetting layers, diamagnetic effects of exciton states provide an effective way to investigate the coupling between two systems with different dimensionality. This article reports for the first time the coupling between a QD and the wetting layer through observation of many-body exciton states with a high excitation power in a magnetic field. Strong diamagnetic shifts in the coupled system are observed with diamagnetic coefficients up to 93.0 and −51.7 μeV/T$^2$ for positive and negative diamagnetic shifts, respectively. Recombinations between electrons in either the QD or the wetting layer and holes in the QD are clearly demonstrated by analyzing photoluminescence (PL) spectra in a magnetic field with different configurations.

The QD wafer was grown by molecular beam epitaxy on a [100]-oriented intrinsic GaAs substrate. The layout of the sample structure from the substrate to surface consists of a distributed Bragg reflector composed of 13 pairs of Al$_{0.94}$Ga$_{0.06}$As/GaAs (67/71 nm), a 200 nm intrinsic GaAs buffer layer, a Si δ-doped GaAs layer with a doping density $N_d = 5 \times 10^{12}$ cm$^{-2}$ that forms a two-dimensional electron gas (2DEG) region, a 50 nm intrinsic GaAs tunneling layer, and an InAs QD layer with a dot density less than $1 \times 10^9$ cm$^{-2}$, as shown in Fig. 1(a). A 200 nm intrinsic GaAs barrier layer was grown on top. During the growth of the QD, the substrate was not rotated to obtain a graded QD density. This technique enabled us to find a suitable region where the dot density is low enough to address single QDs. Pyramidal QDs with a diameter of 20 nm and height of 6 nm having the Zincblende structure were observed by high-resolution cross-section transmission electron microscopy, as reported previously [42].

To fabricate Schottky diode devices, first mesa isolation trenches were patterned on the wafer using photolithography. The ohmic contact for 2DEG in the n-type region was formed by evaporating a AuGeNi layer followed by alloy annealing. Semitransparent Ti with a thickness of 10 nm was evaporated to form the Schottky contact, followed by the deposition of an Al mask with different apertures of 1–2 μm diameter for optically addressing single QDs. Finally, Cr–Au bond pads were evaporated on top of ohmic contacts and Al masks, as shown in Fig. 1(b).

The device was mounted on an *xyz* piezoelectric stage in a helium gas exchange cryostat equipped with a superconducting magnet. Measurements were carried out at 4.2 K using a magnetic field up to 9 T in Faraday geometry and 4 T in Voigt geometry. The bias voltage was applied to Schottky contacts with a source-measurement unit. Figure 1(c) shows the typical *I–V* curve of the Schottky diode, indicating a rectifying behavior. To perform micro-PL measurements with a confocal microscope, a semiconductor laser with a wavelength of 650 nm was used as the pumping source and was focused on one of the apertures by using a microscope objective with a large numerical aperture (NA) of 0.83. The pumping power was calibrated on top of the sample surface. The PL from single QDs was collected with the same objective, coupled to a multimode fiber, and then dispersed through a 0.55 m spectrometer. The spectrum was acquired using a liquid nitrogen cooled charge coupled device camera with a spectral resolution of 60 μeV.

Figure 2 shows the PL spectra at different pumping powers for a single QD in a Schottky diode structure without an external bias. At a low excitation power of 0.71 μW, only a single peak from the doubly negatively charged exciton ($X^{2-}$) can be observed. More exciton states, such as the singly negatively charged exciton $X^-$ and neutral exciton $X^0$, are observed with the increase of excitation power. The second peak of $X^{2-}$ at the low-energy side is due to the exchange-split final states, singlet and triplet states [27]. The main PL peaks are negatively charged, which is due to the built-in electric field in the Schottky diode and the different injection rates for electrons and holes with non-resonant pumping [14]. The details of peak assignation as a function of the external bias are shown in Fig. S1 in the Electronic Supplementary Material (ESM). With an excitation power of 11.85 μW, a number of small peaks with relatively low intensities appear, as shown with empty circles and triangles in the top spectrum in Fig. 2. These peaks have been neglected in previous studies because these could likely be attributed to neighboring QDs when using a high pumping power and are very difficult to be identified clearly. Here, we focus on these peaks and their magnetic properties.

When a magnetic field is applied along the growth direction of the QDs, namely in Faraday geometry, the PL peaks arising from different charged states in a QD split linearly with increasing magnetic field as a result of the Zeeman effect [14, 27, 43]. Zeeman splitting has been observed for different charged exciton states ($X^+$, $X^0$, $X^-$, and $X^{2-}$), as shown in Fig. S2 in the ESM. The binding energy between $X^-$ and $X^{2-}$ is only approximately 370 μeV, which is considerably lower than that between $X^+$ and $X^0$ or between $X^-$ and $X^0$. The third electron in $X^{2-}$ occupies the p-shell and has less wavefunction overlap with the electrons and holes in the s-shell, resulting in a smaller Coulomb interaction [14]. Because of the larger injection rate for electrons than that for holes in non-resonant pumping, the exciton states are more negatively charged with increasing pumping power. In addition, the charging states become more negatively charged when the bias voltage is changed from −0.5 to 0.5 V, which is due to reduced electron tunneling rates at a positive bias voltage.

Figure 3 shows PL spectra as a function of magnetic field with different applied bias voltages at pumping power of 7.11 μW and 11.85 μW. At 7.11 μW, only $X^0$, $X^-$, and $X^{2-}$ can be identified in Fig. 3(a) at a bias voltage of −0.5 V. With the increase of bias voltage, $X^0$ disappears and $X^{2-}$ becomes stronger (Figs. 3 (a)–3(e)). The excitonic peaks of $X^0$, $X^-$, and $X^{2-}$ show a clear Zeeman splitting and a diamagnetic shift as a function of magnetic field, as reported earlier [14]. Two branches of $X^{2-}$ with singlet and triplet final states show opposite diamagnetic shifts. A positive (negative) diamagnetic coefficient of 7.9 μeV/T$^2$ (−14.7 μeV/T$^2$) is observed for $X^{2-}$ with the triplet (singlet) final state after recombination [27]. Coupling between $X^{3-}$ and the wetting layer shows anti-crossing with increasing magnetic field as shown in Fig. S3 in the ESM, similar to previous observations [27]. Surprisingly, with the increase of the bias voltage, a bunch of small peaks appear, which show a large negative diamagnetic shift and a small g-factor. The enhanced intensities of these small peaks with increasing bias voltage indicate that the exciton states are more negatively charged in the coupled system. Similarly, more peaks can be obtained with increasing pumping power up to 11.85 μW as shown in Figs. 3(f)–3(j). Furthermore, a few peaks with a strong positive diamagnetic shift can be resolved in addition to peaks with a negative diamagnetic shift. All peaks with negative and positive diamagnetic shifts are labeled with red circles and green triangles, respectively, in Fig. 3(j); these peaks are located at the same energy positions as marked on the top trace in Fig. 2.

Since the applied magnetic field acts as a perturbation in the weak-field limit, the PL peaks shift towards higher energy quadratically as a function of magnetic field ($B$) in Faraday geometry. The energy shift of exciton states can be expressed as $\Delta E = \gamma B^2$, where the diamagnetic coefficient $\gamma = e^2 \ell_\alpha^2 / 8 m_\alpha^*$, where $\alpha = e$ or $h$ denotes an electron or a hole, and $m_\alpha^*$ is the effective mass of the electron or hole, and $\ell_\alpha$ is the lateral expansion of the carrier's wavefunction perpendicular to the magnetic field [44]. For self-assembled QDs grown by using molecular beam epitaxy, the diamagnetic coefficient $\gamma$ is approximately 5–10 μeV/T$^2$ for strongly confined QDs [14, 39, 40, 44], and up to 32.8 μeV/T$^2$ for large elongated QDs [45]. Figure 4 shows the diamagnetic coefficients of all the resolved PL peaks with different excitonic energies. For a particular QD, the diamagnetic coefficients of X$^0$, X$^-$, and X$^{2-}$ in the QD are in the range of 5–8 μeV/T$^2$. However, the maximum value for other peaks is 93.0 μeV/T$^2$, which is close to the bulk limit of 110.2 μeV/T$^2$ [38]. The high values of $\gamma$ indicate a greater expansion of wavefunctions for these exciton states. Since the holes are confined in the dot, we only consider the diamagnetic coefficient of electrons in our system. The wavefunction extent $\ell_e$ is approximately 12–15 nm with an electron effective mass of $m_e^* = 0.05 m_0$, where $m_0$ is the free electron mass. The electron wavefunciton extent is much larger than that of the neutral exciton X$^0$ confined in the 4.0 nm QD with a $\gamma$ of 6.9 μeV/T$^2$. The wavefunction extent of 12–15 nm is similar to the magnetic length 15 nm at 3T [40].

Strong excitation power and positive bias voltages are required to observe the extra PL peaks, which are also the criteria for achieving highly negatively charged exciton states in our structure [14]. With increasing excitation power, optically generated electrons prefer to occupy the ground state (s-shell) and the first excited state (p-shell) of the conduction band of the QD, followed by the continuum state in the wetting layer. The second excited state cannot be filled because the energy levels are much higher than those of the continuum states of the wetting layer for small QDs. However, these PL peaks with large $\gamma$ cannot be ascribed to the recombination of electron–hole pairs in the wetting layer, as the exciton energy of the wetting layer is much larger than that of QDs, and the linewidth is much broader.

We ascribe these PL peaks with large $\gamma$ to Mahan exciton states [21, 46], where the electrons occupying a QD or a wetting layer recombines with holes in the QD only, as shown in Fig. 5. However, it is very complex to assign all the PL peaks to the charging states with the precise number of electrons and holes, as reported in a previous study [13]. Nevertheless, the energy separation between the adjacent peaks is approximately a few milli-electronvolts, which is in the range of the Coulomb energy for different charging states. With these configurations, there are two possible recombination paths between electrons in the wetting layer or in the QD and holes in the QD, resulting in two final states, as shown in Figs. 5(a) and 5(b), respectively. The large wavefunction expansion of the holes in the ground state can also be eliminated because the holes are well confined in small QDs owing to their large effective mass. This can also be confirmed by the fact that the $\gamma$ of singly positively charged exciton $X^+$ is approximately 7.4 µeV/T$^2$, as shown in Fig. S2 in the ESM. Therefore, the large diamagnetic coefficient can only be due to the large wavefunction expansion of the excited electron states that diffuse into the wetting layer, a quasi-two-dimensional layer close to the QDs. Hence, we attribute the PL peaks with strong diamagnetic shifts to the participation of electrons in the wetting layer.

Since diamagnetic shifts of normal exciton states confined in QDs have been reported before [14, 40], we only consider the wavefunction expansion of excited electron states with large diamagnetic shifts to understand the coupling between two systems. For a single quantum well, the diamagnetic shift decreases with decreasing well width when a magnetic field is applied perpendicular to the well plane. However, for very thin quantum wells (less than 5 nm thick) with finite potential confinement, an increase of $\gamma$ has been observed with the decrease of the well width because of the expansion of the wavefunction into the barrier regions [38]. The monolayer-thick wetting layer can be considered as a thin quantum well, as characterized by transmission electron microscopy [34, 35, 42]. In a system with a single QD coupled with a wetting layer, the optically generated electrons using high excitation power occupy either the wetting layer or excited states in the QDs. Figures 6(a) and 6(d) show the schematic wavefunction distributions of electrons in the wetting layer and QD, respectively. The wavefunction extents are scaled to $\ell$ values derived from the observed diamagnetic coefficients. The cross sections of the wavefunction distributions at different viewing angles are shown in Fig. 6. It can be seen that the electron wavefunction in the wetting layer spreads to a much larger extent than that in the QD plane perpendicular to the magnetic field in Faraday geometry (Fig. 6(b)), which results in a large positive diamagnetic shift (denoted by green empty triangles in Fig. 4) for the recombination of electrons in the wetting layer and holes in the QD. Figure 6(e) shows the top view of the electron wavefunction distribution in the QD, which has a much smaller diameter compared with that in the wetting layer. The small wavefunction distribution exhibits regular diamagnetic shifts, as shown by black squares in Fig. 4.

Next, we discuss the origin of negative diamagnetic shifts for other PL peaks at a high excitation power. Since no linewidth broadening is observed with increasing magnetic field, it can be considered that shakeup processes have no contribution to the negative diamagnetic shift [47]. When electrons and holes occupy the quantum dot or the wetting layer with the application of a high pumping power, which is in the regime of weak confinement, the Coulomb energies are equal to or considerably higher than the single-particle energies. In this regime, the diamagnetic coefficient is proportional to the difference of the wavefunction extent between the initial and final states of radiative recombination [41]. Several research groups have observed anomalous negative diamagnetic shifts for negatively charged excitons, which can be explained by different wavefunction expansions of the initial and the final states [40, 41, 48]. The diamagnetic coefficient $\gamma$ is proportional to $\ell_i^2 / m_i^* - \ell_f^2 / m_f^*$, where $\ell_i$ ($\ell_f$) and $m_i^*$ ($m_f^*$) are the wavefunction extent and the effective mass of the exciton complex in the initial (final) state. In the initial state of negatively charged excitons, the electron wavefunctions are confined in the dot because of the strong Coulomb attraction of holes. Following the recombination of an electron and a hole, the wavefunctions of the remaining electrons expand, resulting in a negative diamagnetic shift (paramagnetic effect). For singly negatively charged exciton X$^-$ in a small QD, $\gamma$ values ranging from −5 to −10 μeV/T$^2$ have been reported [40]. However, in our system, the maximum diamagnetic coefficient is −51.7 μeV/T$^2$, which is approximately five times larger than that of X$^-$. The huge expansion can be explained as follows. Initially, electrons are confined in the QD as a result of the attraction of the holes in the QD, as shown in Fig. 6(e). After recombination, the attraction of the holes disappears and one or more electrons spread to the wetting layer with a large expansion. Another possible reason is that the electrons in the wetting layer also expand after the disappearance of the attractive force of holes in the QD. Since there is no confinement in the wetting layer, a large increase of the wavefunction distribution from the QD (Fig. 6 (e)) to the wetting layer (Fig. 6(b)) induces a strong negative diamagnetic shift, which also confirms the coupling between the QD and the wetting layer.

To further verify our model, the same measurements were performed by changing the direction of the applied magnetic field from parallel (Faraday geometry) to perpendicular (Voigt geometry) to the direction of the QD growth. Figure 7 shows the contour plots of the PL spectra as a function of magnetic field aligned at angles (denoted as $\theta$) of 0°, 45°, and 90° to the growth direction. Figures 7(d)–7(f) show the Zeeman splitting and diamagnetic shifts of the high-energy branch from X$^{2-}$ in different magnetic configurations. When $\theta$ increases from 0° to 90°, the Zeeman splitting decreases from 650 to 98 μeV at 4 T owing to the negligible in-plane g-factor of holes in Voigt geometry [49]. The diamagnetic coefficient $\gamma$ of X$^{2-}$ decreases from 7.9 to 2.1 μeV/T$^2$, indicating that the wavefunction extent perpendicular to the growth direction is larger than that parallel to it, as demonstrated in Figs. 6(e) and 6(f). This corresponds well with the truncated pyramidal shape of our QDs [42]. The low $\gamma$ value (2.1 μeV/T$^2$) indicates a small wavefunction expansion along the growth direction. The Zeeman splitting of the peaks with strong positive diamagnetic shifts is not resolvable. The positive diamagnetic coefficient decreases sharply from 90.6 μeV/T$^2$ to nearly zero as $\theta$ increases from 0° to 90°, as shown in Figs. 7(a)–7(c). This behavior further confirms that these peaks originate from the recombination of electrons in the wetting layer and holes in the QD. With the magnetic field perpendicular to the growth direction, the wavefunction expansion along the growth direction is mainly confined in the wetting layer with a small distribution for cyclotron motion, as shown by dashed circles in Fig. 6(c). The reduced diameter of the wavefunction perpendicular to the magnetic field results in a small diamagnetic coefficient, which cannot be resolved in Voigt geometry (as shown in Fig. 7(c)).

Figures 7(g)–7(i) show the contour plots of the magneto-PL spectra of a peak with a negative diamagnetic shift for different magnetic field configurations. With the increase of $\theta$ from 0° to 45°, the diamagnetic coefficient $\gamma$ changes from −51.7 to −26.2 μeV/T$^2$, indicating a reduced wavefunction expansion difference between the initial and the final states along the direction perpendicular to the magnetic field. Surprisingly, a positive value of $\gamma$ (2.5 μeV/T$^2$) is observed for this peak in Voigt geometry, which is similar to the result of X$^{2-}$. It is evident that the final-state wavefunction expansion to the wetting layer does not contribute to the diamagnetic shift in Voigt geometry, which instead is caused by the wavefunction distribution along the growth direction of QDs, as shown in Fig. 6(f).

In the case of the shakeup process or tunneling effect, the coupling between a single QD and a degenerate electron gas normally leads to linewidth broadening at different bias voltages [21, 29, 30]. However, in our coupled system, no linewidth broadening is observed in the PL peaks in the applied voltage range. This could be because only a few electrons in the wetting layer are strongly coupled to the carriers in a single QD through Coulomb interaction. The narrow linewidth indicates a coherent exciton state [50], which has a great potential for application in solid-state quantum information processing using a coupled system.

In summary, we demonstrated a coupled QD –wetting layer system with large diamagnetic coefficients of many-body exciton states by using magneto-photoluminescence spectroscopy. With a low excitation power, different charged states (X$^+$, X$^0$, X$^-$, X$^{2-}$, and X$^{3-}$) were observed with diamagnetic coefficients in the range of 5–10 μeV/T$^2$. With the increase of excitation power, optically generated electrons occupy the excited states of the QDs and wetting layer, forming many-body exciton states through Coulomb interaction. A bunch of extra PL peaks were observed, which were attributable to many-body Mahan exciton states resulting from the recombination of electrons in the QD or wetting layer and holes in the QD. When electrons in the wetting layer recombined with holes in the QD, a large positive diamagnetic coefficient close to that of the bulk material was observed because of the large wavefunction expansion in the wetting layer. In addition, a large negative diamagnetic effect was observed for the recombination of electrons in the QD and holes because of a huge difference of the wavefunction extents between the initial and final states. The observation of large diamagnetic coefficients of strongly coupled many-body states provides a direct evidence of a coupled structure comprising zero- and two-dimensional systems. This opens a new way to investigate many-body physics, such as the Kondo physics and Fano effect, and to implement solid-state quantum information processing in the future.

## Acknowledgements

This work was supported by the National Basic Research Program of China (Nos. 2013CB328706 and 2014CB921003), the National Natural Science Foundation of China (Nos. 91436101, 11174356, and 61275060), the Strategic Priority Research Program of the Chinese Academy of Sciences (No. XDB07030200), and the 100 Talents Program of Chinese Academy of Sciences. We thank Jean-Pierre Leburton for helpful discussions.

**Figures and captions**

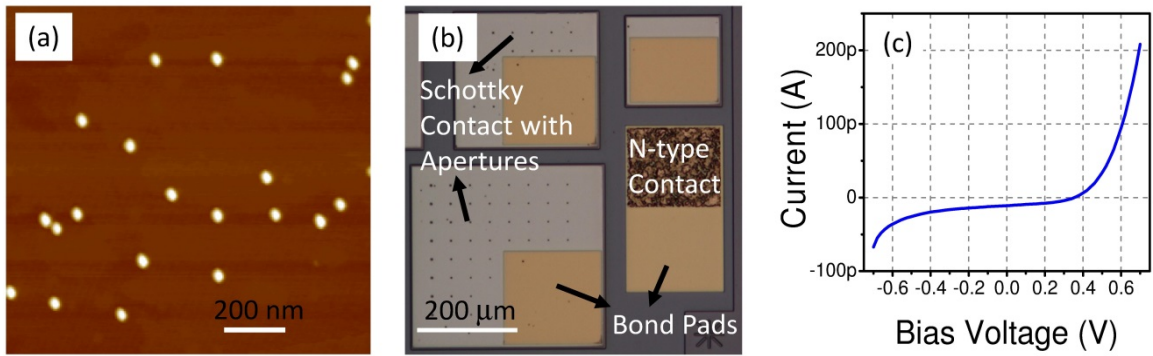

**Figure 1** (a) Atomic force microscopy image of a typical InAs quantum dot wafer with a dot density of $10^9 \cdot cm^{-2}$. (b) Optical microscopy image of the Schottky diode structure with arrays of small apertures to address single quantum dots. (c) A typical *I–V* curve of the Schottky diode with a mesa area of 400 μm × 400 μm.

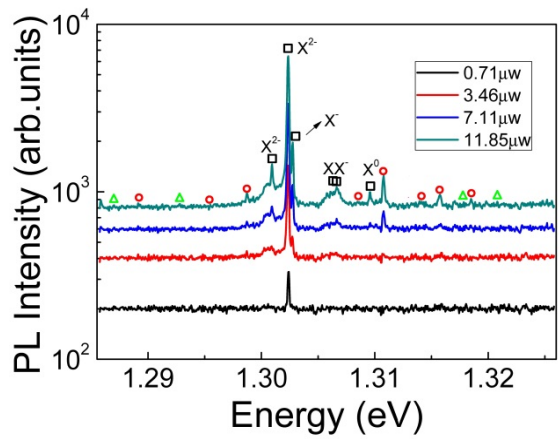

**Figure 2** PL spectra with different pumping power from 0.71 to 11.85 μW. The lines are shifted for clarity. Black squares denote different exciton states, labeled as the neutral exciton ($X^0$), singly negatively charged double exciton ($XX^-$), singly negatively charged exciton ($X^-$), and doubly negatively charged exciton ($X^{2-}$). With a high excitation power of 11.85 μW, a bunch of small peaks appear as shown in the top trace, which are marked with red circles and green triangles. The circles and triangles denote the PL peaks exhibiting negative and positive diamagnetic shifts in the magnetic field, respectively.

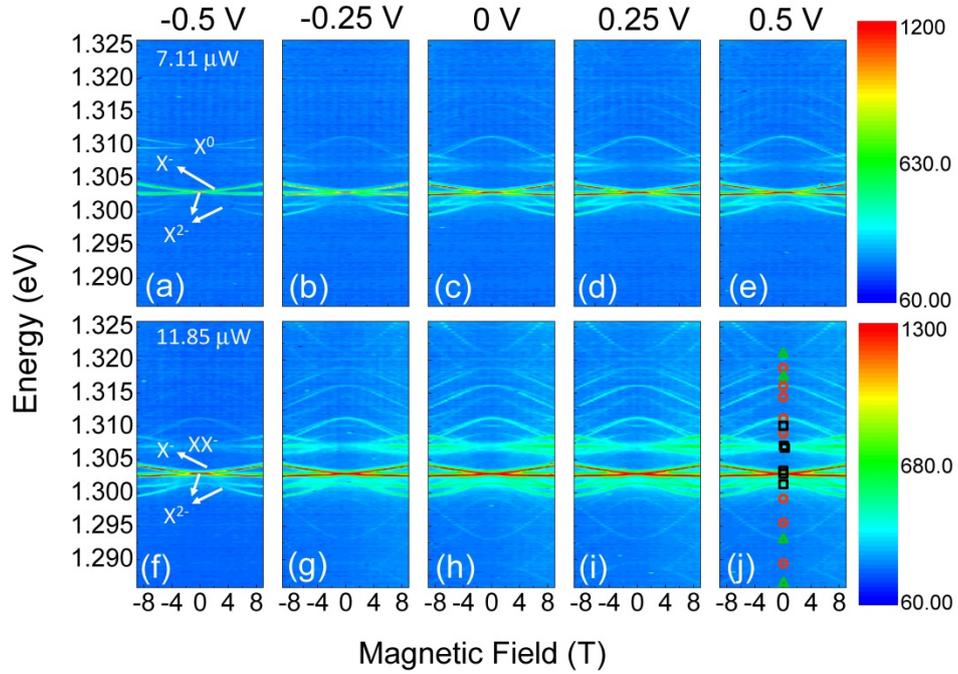

**Figure 3** The contour plots of PL spectra as a function of magnetic field with different bias voltages at a pumping power of 7.11 μW ((a)–(e)) and 11.85 μW ((f)–(j)). The bias voltages are labeled on the top of each column. With a pumping power of 7.11 μW, the neutral exciton $X^0$ can only be observed under a bias voltage of −0.5 V. The intensities of $X^{2-}$ are enhanced when the bias voltages are changed from −0.5 to 0.5 V because of the reduced electron tunneling rate at positive bias voltages. When the pumping power is 11.85 μW, the dot is more negatively charged, as discussed in the text. Additional small peaks appear with strong diamagnetic shifts in the magnetic field. These peaks are marked with green triangles and red circles in (j), as labeled on the top trace in Fig. 1. In addition, singly negatively charged double exciton peaks can be resolved at a high pumping power in (f)–(j).

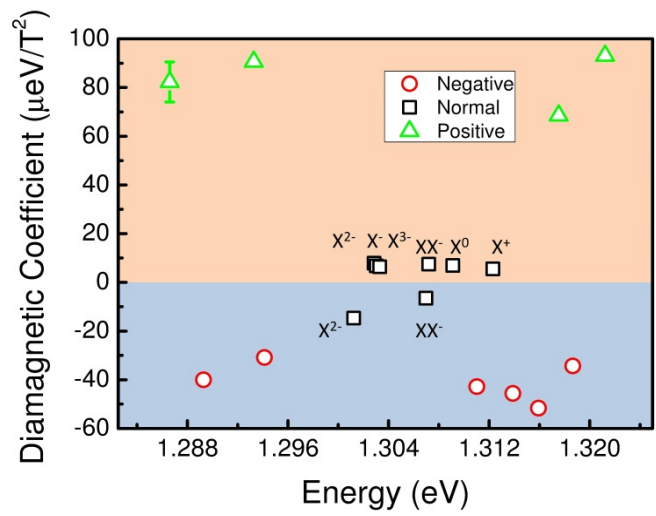

**Figure 4** Diamagnetic coefficients of different charge states in a coupled system at a pumping power of 11.85 μW and a bias voltage of +0.5 V. The values for $X^0$ and $X^+$ are extracted from the spectra corresponding to a pumping power of 5.93 μW and a bias voltage of −0.5 V, as shown in Fig. S2 in the ESM. A typical error bar for a PL peak is shown at 1.2866 eV.

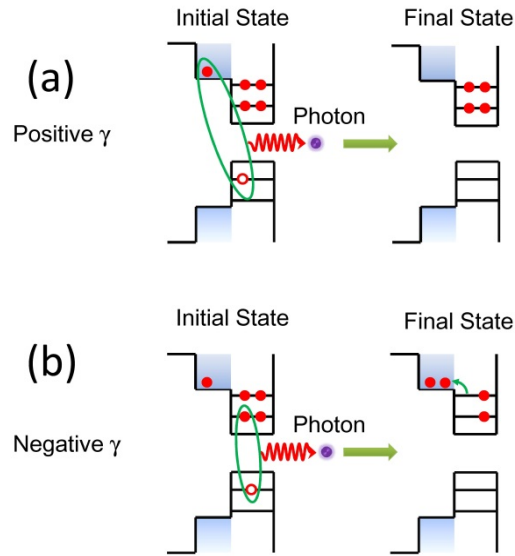

**Figure 5** Schematic energy diagrams of the initial and final states of negatively charged exciton states in a coupled QD–wetting layer system. (a) The electrons in the wetting layer recombine with holes in the QD, resulting in large positive diamagnetic shifts. (b) The electrons recombine with the holes in the QD, inducing large wavefunction distribution differences between the initial and final states. Owing to the absence of attraction of the holes in the QD after recombination, the electrons in the wetting layer spread in the final states and electrons in the QD may be released to the wetting layer and then expand, as indicated by the green arrow. This expansion induces a larger wavefunction extent of the final states than that of the initial states of the system, resulting in a negative diamagnetic effect.

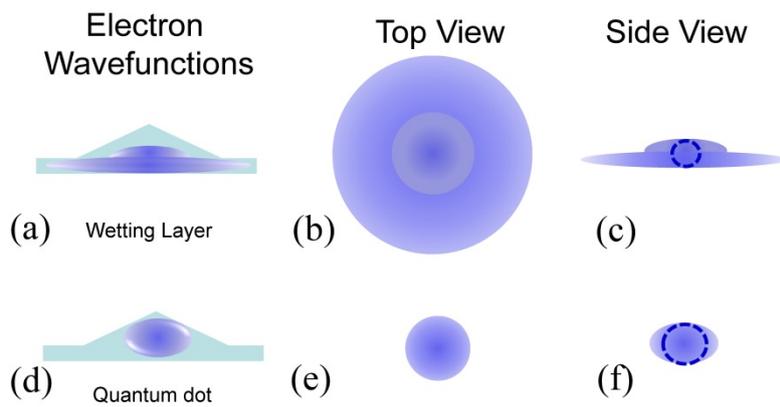

**Figure 6** Schematic of electronic wavefunctions in a coupled QD–wetting layer system from different viewing angles. Wavefunction distributions of electrons occupying the wetting layer are shown in (a)–(c) and those in the QD in (d)–(f). The sizes of wavefunciton extents are scaled to $\ell$ extracted from diamagnetic coefficients. The lateral distribution of electrons in the wetting layer is much larger than that along the growth direction, as shown by the dashed circle in (c). This induces a large positive diamagnetic shift on application of a magnetic field in Faraday geometry, but nearly zero shift in Voigt configuration. However, the difference between the wavefunction distributions in the QDs is much smaller as shown in (e) and (f).

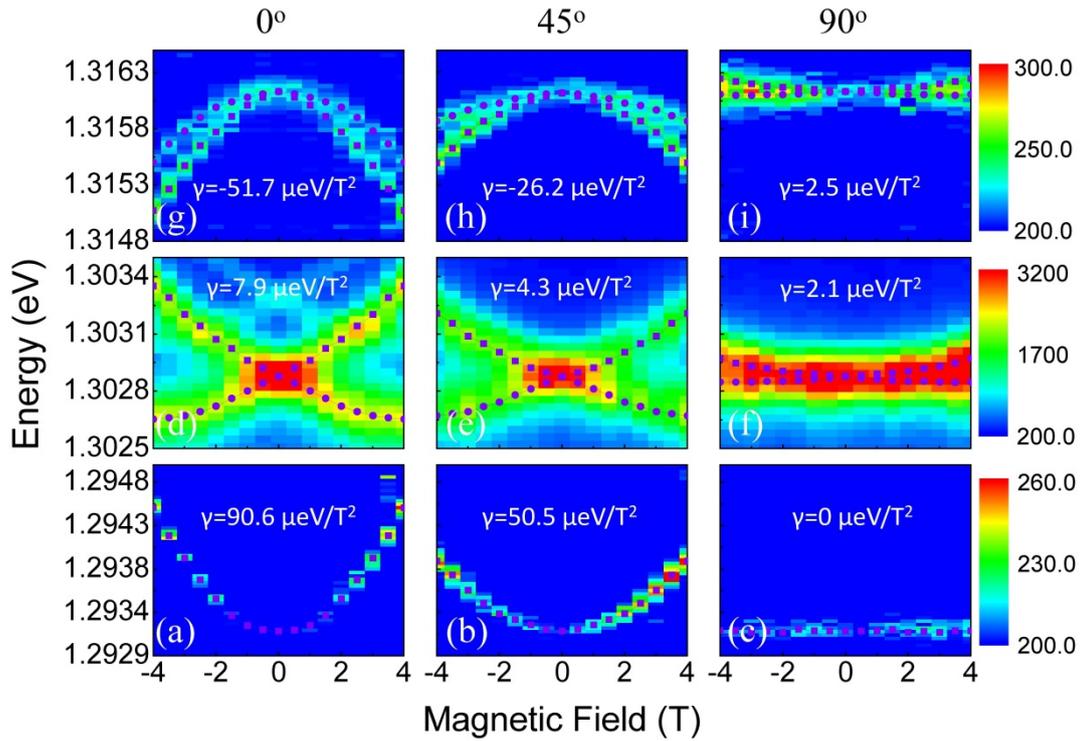

**Figure 7** PL spectra as a function of applied magnetic field in different configurations. The angle $\theta$ between the applied magnetic field and growth direction is labeled on top of each column. (a)–(c) show diamagnetic shifts in different configurations for a typical PL peak, which has a large positive diamagnetic shift in Faraday geometry. (d)–(f) are the results for the upper branch of $X^{2-}$ and (g)–(i) show a typical PL peak with a large negative diamagnetic shift. The calculated diamagnetic coefficients are shown in each panel.